# Understanding of Genetic Code Degeneracy and New Way of Classifying of Protein Family: A Mathematical Approach

Jayanta Kumar Das[1], Atrayee Majumder[2], Pabitra Pal Choudhury[3]

Applied Statistics Unit
Indian Statistical Institute
Kolkata-700108, India

[1]dasjayantakumar89@gmail.com , atu.tua@gmail.com[2], pabitrapalchoudhury@gmail.com[3]

*Abstract—* **The genetic code is the set of rules by which information encoded in genetic material (DNA or RNA sequences) is translated into proteins (amino acid sequences) by living cells. The code defines a mapping between tri-nucleotide sequences, called codons, and amino acids. Since there are 20 amino acids and 64 possible tri-nucleotide sequences, more than one among these 64 triplets can code for a single amino acid which incorporates the problem of degeneracy. This manuscript explains the underlying logic of degeneracy of genetic code based on a mathematical point of view using a parameter named "Impression". Classification of protein family is also a long standing problem in the field of Bio-chemistry and Genomics. Proteins belonging to a particular class have some similar bio-chemical properties which are of utmost importance for new drug design. Using the same parameter "Impression" and using graph theoretic properties we have also devised a new way of classifying a protein family.**

*Index Terms*- Encoding, Impression, Order pair Graph, Degeneracy, Classification etc.

## I. Introduction

Biological functionality of every living organism is regulated by proteins [1] which, in turn, can be viewed as sequences of amino acids. There are many methods for information processing that can be used for the analysis of proteins (sequence of amino acids) [2, 3]. There are 25 amino acids among which 20 are standard and 5 are non-standard amino acids. All of them can be represented by their respective unique symbols in ternary notation following an encoding scheme. The main problem is how to make the digital coding of amino acids for better distinguishability which can be done in various ways [4]. Here encoding is done based on molecular weight of amino acid. The question may arise why as such encoding schemes are required. From biochemical point of view, amino acids have some properties (polar, non-polar, charged etc.) with the help of which we can categorize them into different groups. But when we want to study these groups from a mathematical point of view we need to define parameters which can uniquely represent these 25 amino acids without any ambiguity. Therefore we have to transform each amino acid to some other notable symbols so that we can formulate a new parameter to study the properties of the group mathematically. For this reason, encoding or transformation of amino acids into ternary numbers get suitable notation and best fitted without loss of generality.

The code in which genetic instructions are written, using an alphabet based on the four bases in DNA and RNA: adenine (A), cytosine (C), guanine (G), and thymine (T) (for DNA) or uracil (U) (for RNA). Each triplet of bases indicates a particular kind of amino acid which is to be synthesized. Since there are 20 amino acids and 64 possible triplets, more than one triplet can code for a particular amino acid [5]. The code is non-overlapping; the triplets are read end-to-end in sequence (e.g. UUU = phenylalanine, UUA = leucine, CCU = proline); and there are three triplets not translated into amino acid, indicating chain termination called stop codon. The code is universal and applies to all species with some exceptions. Theoretically, there are $4^3$ = 64 different codon combinations possible with a triplet codon of three nucleotides. In reality, all 64 codons of the standard genetic code are assigned for either amino acids or stop signals during translation. If, for example, an RNA sequence, UUUAAACCC is considered and the reading-frame starts with the first U (by convention, 5' to 3'), there are three codons, namely, UUU, AAA and CCC, each of which specifies one amino acid. Fig. 1 shows codons specifying each of the 20 standard amino acids involved in translation.

Fig 1. Codon Table

Using Hasse diagram an attempt has been made to correlate different hydrophobicity's of amino acids and their respective codons [6].

There are numerous types of protein families and functionality of each protein family is quite diverse. There is no well-defined mathematics available which can address the existence of different classes in a protein family. To find the mathematics behind a particular class is quite a hard problem. Therefore it is required for establishing a new classification methodology from mathematical point of view and then correlating the biological properties with the mathematical properties (if any) for a given class. This would help to understand the protein classes more elaborately such that chemical/biological properties of each class can be correlated with the mathematical properties of the class and hence this will facilitate new drug design.

Once encoding is done, we have defined one mathematical parameter "Impression" of amino acid on ternary numbers which is in triplet form. Based on this parameter first we explain the degeneracy of codon table and secondly using graph theoretic model we classify the proteins the iron protein family [7] of existing classes where their classification is done based on bio-chemical point of view.

The paper is organized as follows: Section II discussed the encoding of amino acid into ternary numbers, introduction to "Impression" parameter and Graph building process using "Impression". Section III deals with results and discussion of degeneracy of codon table and classification of Iron protein family followed by conclusion of this manuscript.

## II. METHODS AND MATERIALS

### A. Encoding of Amino Acids:

An amino acid can be replaced by three ternary symbols $X_1, X_2$ and $X_3$ representing a 1-variable ternary number/symbols where $X_1, X_2, X_3 \in \{0,1,2\}$. Using these three ternary symbols/numbers there are 27 combinations of ternary numbers. Proteins are composed of just combination of 20 conventional and 5 non-conventional different amino acids. We can fit our 25 amino acids and we are not leaving any ternary number blank, by putting smallest ternary number as gap (" ") and highest ternary number as unknown ("X") amino acids.

But big question is in which ordering the encoding of amino acids into ternary numbers will be. Varieties of encoding schemes can be done by ordering the amino acids like hydrophobic indexing, molecular weight, natural abundance etc. A particular encoding scheme may resolve a specific genomics problem. Here we have organized the amino acids based on their molecular weight which is well enough to resolve our problems addressing the degeneracy of codon table and classification of a particular protein family. Calculations of molecular weight in gm/mol of amino acids are shown in TABLE I. By ordering them in ascending order amino acids to corresponding ternary numbers in order is shown in TABLE II.

TABLE I. AMINO ACIDS AND THEIR MOLECULAR WEIGHT

| Amino Acids | Molecular weight (g/mol) | Amino Acids | Molecular Weight (g/mol) |
|---|---|---|---|
| Alanine(A) | 89.0935 | Proline(P) | 115.1310 |
| Cysteine(C) | 121.1590 | Glutamine(Q) | 146.1451 |
| Aspartate(D) | 133.1032 | Arginine(R) | 174.2017 |
| Glutamate(E) | 147.1299 | Serine(S) | 105.0930 |
| Phenylalanine(F) | 165.1900 | Threonine(T) | 119.1197 |
| Glycine(G) | 75.0669 | Selenocysteine(U) | 168.0500 |
| Histidine(H) | 155.1552 | Valine(V) | 117.1469 |
| Isoleucine(I) | 131.1736 | Tryptophan(W) | 204.2262 |
| Lysine(K) | 146.1882 | Tyrosine(Y) | 181.1894 |
| Leucine(L) | 131.1736 | N-Formylmethionine (fMet) | 177.2200 |
| Methionine(M) | 149.2124 | Hydroxyproline (Hyp) | 131.1300 |
| Asparagine(N) | 132.1184 | Hydroxylysine (Hyl) | 162.1870 |
| Pyrrolysine(O) | 255.3100 | | |

TABLE II. ENCODING OF AMINO ACIDS INTO TERNARY NUMBERS

| Type | Code | | | | | | | |
|---|---|---|---|---|---|---|---|---|
| Character | GAP | G | A | S | P | V | T | C | Hyp |
| Decimal | 0 | 1 | 2 | 3 | 4 | 5 | 6 | 7 | 8 |
| Ternary | 000 | 001 | 002 | 010 | 011 | 012 | 020 | 021 | 022 |
| Character | L | I | N | D | Q | K | E | M | H |
| Decimal | 9 | 10 | 11 | 12 | 13 | 14 | 15 | 16 | 17 |
| Ternary | 100 | 101 | 102 | 110 | 111 | 112 | 120 | 121 | 122 |
| Character | Hyl | F | U | R | fMet | Y | W | O | X |
| Decimal | 18 | 19 | 20 | 21 | 22 | 23 | 24 | 25 | 26 |
| Ternary | 200 | 201 | 202 | 210 | 211 | 212 | 220 | 221 | 222 |

### B. Impression of Amino Acids:

Once ternary numbers are assigned to amino acids, we can think of an amino acid as three ternary symbols $X_1 X_2 X_3$ where $X_1, X_2, X_3 \in \{0,1,2\}$. Impression of amino acid denoted by IP and is defining the triplet ($I_1, I_2, I_3$) as follows:-

$$IP(X_1, X_2, X_3) = (I_1, I_2, I_3) \cdots (1)$$

Summoned of the triplet is the Total Impression (TIP) as follows-

$$TIP = I_1 + I_2 + I_3 \cdots (2)$$

where $I_1 = Sym(X_1, X_2), I_2 = Sym(X_2, X_3),$
$I_3 = Sym(X_3, X_1)$ and $Sym(X_i, X_j) = |X_i - X_j|$

(Following Abbreviations are used throughout this paper: **TS-**Ternary Symbols, **AA-**Amino Acid, **IP-**Impression, and **TIP-**Total Impression)

Using equation (1) and (2) calculated IP and TIP values respectively for 27 ternary numbers including 25 amino acids and other two for GAP and X are shown in TABLE III. It is clearly observed from Table 3 that we have 27 ternary numbers which are mapped into 10 IP values (a group of triplet form). Again 10 IP values are broadly categorized into 3 groups and several subgroups but TIP value is same within a group.

- **First group**- there are three ternary numbers mapped into one IP value (0, 0, 0) and only one TIP value which is 0 and there is no sub groups.
- **Second group**-there are twelve ternary numbers mapped into three subgroups i.e. IP values are (0, 1, 1), (1, 0, 1) & (1, 1, 0), each with four ternary numbers and all their TIP value is 2.
- **Third group**-there are twelve ternary numbers mapped into six subgroups i.e. IP values are (0, 2, 2), (2, 0, 2), (2, 2, 0), (1, 2, 1), (1, 1, 2) & (2, 1, 1) each with two ternary numbers and all their TIP value is 4.

TABLE III. IMPRESSION TABLE

| Groups | TS of AA | IP $(I_1,I_2,I_3)$ | TIP $(I_1+I_2+I_3)$ |
|---|---|---|---|
| 1st group | GAP-000 Q-111 X-222 | (0, 0, 0) | 0 |
| 2nd group | G-001 D-110 K-112 O-221 | (0, 1, 1) | 2 |
| | L-100 P-011 fMet-211 H-122 | (1, 0, 1) | |
| | S-010 I-101 M-121 Y-212 | (1, 1, 0) | |
| 3rd group | A-002 W-220 | (0, 2, 2) | 4 |
| | Hyl-200 Hyp-022 | (2, 0, 2) | |
| | T-020 U-202 | (2, 2, 0) | |
| | E-120 N-102 | (1, 2, 1) | |
| | V-012 R-210 | (1, 1, 2) | |
| | F-201 C-021 | (2, 1, 1) | |

*C. Ordered triplet weighted directed graph using Impression*

A directed graph (G): = {V, E} consists of a set of vertices; where V= {$V_1$, $V_2$… $V_N$} are IP values in triplet form (representing an amino acid) and E= {$E_1$, $E_2$… $E_M$} set of directed edges among the vertices. There exists 10 unique IP values; if we draw a graph using the IP values maximum number of vertices will be 10. Graph is called order triplet directed graph as it is applied on sequence on amino acid where amino acid is IP i.e. in triplet form and two consecutive triplets in order implies a directed edge. For example, let amino acid sequence DAAQHDHD in order and corresponding IP (in triplet form) values are (0, 1, 1), (0, 2, 2), (0, 2, 2), (0, 0, 0), (1, 0, 1), (0, 1, 1), (1, 0, 1) & (0, 1, 1) from Table III. Now, one can draw an order triplet directed graph starting from the vertex (0, 1, 1) and an directed edge vertex (0, 1, 1) to next vertex (0, 2, 2) in order, then (0, 2, 2) to (0, 2, 2) i.e. self-loop in order and so on. Weight of an edge ($E_I$) between two vertices $V_J$ and $V_K$ is denoted as $WE_I$ and is defined as average TIP value of the two terminal vertices i.e.
$$WE_I = \frac{1}{2}\{TIP(V_I) + TIP(V_J)\}.$$ Corresponding graph of the order sequences DAAQHDHD is shown in the Fig 2.

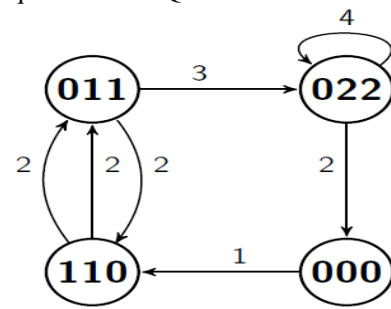

Fig 2. Order Triplet Weighted Directed Graph for the Amino Acid Sequence DAAQHDHD

III. RESULTS AND DISCUSSIONS

Following TABLE IV shows the alternative codon Table of Fig. 1 for 64 codons. Here codon table is organized based on Amino Acid (representing one letter symbol) and corresponding Ternary Symbol, Impression, Total Impression are also shown in this table for each codon.

On closer look in the Table III, one can think of table as 16 groups, each group representing four codons, where each codon is a single amino acid. If 16 groups= {G1, G2, G3, G4}×{G1, G2, G3, G4}, then G1×G1 = {XYZ}; where XYZ {(UUU), (UUC), (UUA), (UUG)} and each representing an Amino Acid.

Total sixteen groups are divided into two types of groups **A) Symmetric Group break** and **B) Asymmetric group break.** Further two types of symmetric group is there, one symmetric group is in the ratio 4:0 i.e. all four codons are representing same ternary symbol and other symmetric group is in the ratio 2:2 i.e. two codons are one type similarity of ternary symbol and other two codons another type similarity of ternary symbol. But in asymmetric group break, ratio is different either 3:1 or 2:1:1. Therefore total three cases, two cases for symmetric group break and one case for asymmetric group break from which we may infer the following result which clearly explains the degeneracy of codon table.

*A. Symmetric Group Break:*

- **Case 1 (4:0 break): TS** are same for all 4 codon and corresponding **IP** is same.

Total 8 groups= {{G2, G1}, {G4, G1}, {G1, G2}, {G2, G2}, {G3, G2}, {G4, G2}, {G2, G4}, {G4, G4}}

- **Case 2 (2:2 break ): TS** are same for 2 codons in two sub groups and corresponding **IP** is different.

TABLE IV. CODON TABLE IN TERMS OF IMPRESSION VALUE

|  | U (G1) | | | C(G2) | | | A(G3) | | | G(G4) | | | |
|---|---|---|---|---|---|---|---|---|---|---|---|---|---|
|  | AA | TS | IP ($I_1, I_2, I_3$) | AA | TS | IP ($I_1, I_2, I_3$) | AA | TS | IP ($I_1, I_2, I_3$) | AA | TS | IP ($I_1, I_2, I_3$) | |
| U (G1) | F F | 201 201 | (2, 1, 1) | S S S S | 010 010 010 010 | (1, 1, 0) | Y Y Stop Stop | 212 212 | (1, 1, 0) | C C Stop W | 021 021 220 | (2, 1, 1) (0, 2, 2) | U C A G |
| U (G1) | L L | 100 100 | (1, 0, 1) |  |  |  |  |  |  |  |  |  |  |
| C (G2) | L L L L | 100 100 100 100 | (1, 0, 1) | P P P P | 011 011 011 011 | (1, 0, 1) | H H | 122 122 | (1, 0, 1) | R R R R | 210 210 210 210 | (1, 1, 2) | U C A G |
| C (G2) |  |  |  |  |  |  | Q Q | 111 111 | (0, 0, 0) |  |  |  |  |
| A (G3) | I I I | 101 101 101 | (1, 1, 0) | T T T T | 020 020 020 020 | (2, 2, 0) | N N | 102 102 | (1, 2, 1) | S S | 010 010 | (1, 1, 0) | U C A G |
| A (G3) | M | 121 | (1, 1, 0) |  |  |  | K K | 112 112 | (0, 1, 1) | R R | 210 210 | (1, 1, 2) |  |
| G (G4) | V V V V | 012 012 012 012 | (1, 1, 2) | A A A A | 002 002 002 002 | (0, 2, 2) | D D | 110 110 | (0, 1, 1) | G G G G | 001 001 001 001 | (0, 1, 1) | U C A G |
| G (G4) |  |  |  |  |  |  | E E | 120 120 | (1, 2, 1) |  |  |  |  |

- Total 6 groups= {{G1, G1}, {G1, G3}, {G2, G3}, {G3, G3}, {G4, G3}, {G3, G4}}

*B. Asymmetric Group Break:*

- **Case-3 (3:1 break and 2:1:1 break): TS** are different for some codon but the corresponding **TIP** is same for a group and is different for different groups. Total 2 groups ={G3, G1} where **TIP**=2 and {G, G4} where **TIP**=4

To get the ratio 3:1 and 2:1:1, Let x=number of sub groups in a particular group.

Asymmetric ratio will be: **(4-TIP/2): {(x-TIP/2): (x-TIP/2)…m times}** such that m*(x-TIP/2) = TIP/2.
For the group {G3,G1}, TIP=2, x=2 and m=1, therefore (4-2/2): {(2-2/2)} = 3:1
For the group {G1,G4}, TIP=4, x=3 and m=2, therefore (4-4/2): {(3-4/2): (3-4/2)} = 2:1:1

Our approach of classifying a protein family targets the already given classification of the dataset based on some existing bio-chemical properties. For the classification of different proteins sequences, Iron protein family is taken from [7] shown in Fig. 3 consisting of 72 protein sequences. Now for each of the 72 sequences (numbers from top to bottom i.e. 1, 2, 3…72 from Fig. 3) we draw a directed graph using the impression values of the amino acids (discussed in section II (C)). In Fig. 3, each protein (amino acid sequence) showing is partial; one can find the complete sequence in [7].

There are various graph based properties like number of vertices, number of edges, number of variable length cycles, weights etc. which can be considered to draw conclusion of its behavior from a directed graph. Based on the length of any cycles, number of a particular length cycle and presence of unique cycles we have classified the protein sequences. The following table V is showing the node/vertex assignments for different impression values. A cycle 1-5-3-1 means a cycle involving vertexes/nodes (0, 0, 0) to (1, 1, 0) to (0, 2, 2) to (0, 0, 0). Our classification result is based upon various 3 length cycles. Table VI is showing the match between existing and resulted classification. The class named Dsr contains the protein sequences as numbers 1, 2, 3, 4. The common unique cycle among those proteins is 1-5-2-1. As a result of experiment it can be seen that some proteins does not pose similarity according to the existing classes. So they are clubbed together with the existing one. For example, an existing class IIIc contains protein (sequence number 38) which contains unique 11 cycles to identify this protein uniquely as noted in third column of table VI. Further IIIc has similarity with the members of the groups Fsr-C, IIId, IIIb, Ia with unique cycle 2-10-5-2 which is common to all Fsr-C, IIId, IIIb, Ia. Note that Fsr-C contains the proteins as represented by the sequence numbers 6, 8, 10, 14, 15, 21. Similarly we find other proteins in other subgroups IIId, IIIb, Ia.

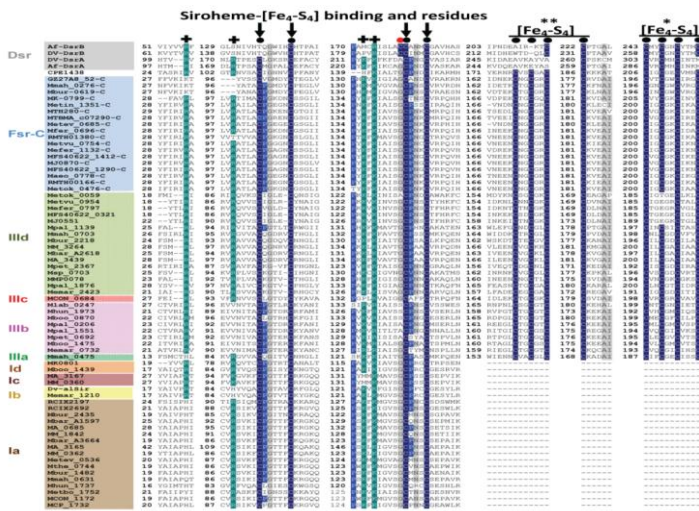

Fig. 3. Iron Protein Family

TABLE V: NODE NUMBERS AND THEIR IMPRESSION VALUE

| Node Number | Impression in Triplet form | Node Number | Impression in Triplet form |
|---|---|---|---|
| 1 | (0, 0, 0) | 6 | (1, 1, 2) |
| 2 | (0, 1, 1) | 7 | (1, 2, 1) |
| 3 | (0, 2, 2) | 8 | (2, 0, 2) |
| 4 | (1, 0, 1) | 9 | (2, 1, 1) |
| 5 | (1, 1, 0) | 10 | (2, 2, 2) |

TABLE V: CLASSIFICATIONS MADE ON UNIQUE CYCLES OF LENGTH 3 PRESENT IN SAME CLASS OF PROTEINS

| Class name | Protein (sequence number) | Unique cycles | Extra protein (sequence number) | Cycles corresponding to extra protein |
|---|---|---|---|---|
| Dsr | {1, 2, 3, 4} | 1-5-2-1 | | |
| NA | {5} | 1-4-5-1<br>1-5-7-1<br>6-10-7-6 | Fsr-C = {7, 9, 11, 12, 13, 16, 17, 18, 19, 20, 22}<br>IIId ={ 27, 37}<br>IIIb = { 39,41 }<br>Ia= {54 } | 2-3-9-2<br>2-10-4-2<br>2-5-10-2<br>3-9-4-3<br>3-9-5-3<br>3-9-6-3<br>4-5-10-4 |
| IIIc | {38} | 1-2-3-1<br>1-5-3-1<br>1-6-3-1<br>1-7-3-1<br>1-9-3-1<br>1-9-4-1<br>1-9-5-1<br>1-5-9-1<br>1-9-6-1<br>1-6-9-1<br>3-9-10-3 | Fsr-C = { 6, 8, 10, 14, 15, 21}<br>IIId = { 28, 30, 31, 32, 33, 34, 35, 36}<br>IIIb = { 40, 42, 43, 44, 45, 46}<br>Ia = {55, 57, 58, 59, 60, 61, 62, 63, 64, 65, 66, 67, 68, 69, 70, 71, 72} | 2-10-5-2 |
| IIIa | { 47 } | 1-7-2-1<br>1-4-9-1<br>7-9-10-7 | IIId = { 23, 24, 25, 26, 29 }<br>Ia = { 56} | 2-3-6-2<br>2-7-3-2<br>2-6-9-2<br>3-6-4-3<br>3-4-7-3<br>3-6-5-3<br>3-5-7-3<br>3-6-7-3<br>4-6-9-4<br>5-6-9-5 |
| Id | { 48, 49 } | Cycles present but greater than length 3 | | |
| Ic | {50, 51} | 1-6-7-1<br>2-10-9-2<br>5-10-9-5<br>6-10-9-6 | | |
| Ib | {52, 53 } | 1-3-2-1<br>1-3-4-1<br>1-3-6-1 | | |

## IV. CONCLUSION

Unique encoding problem of amino acids is a long standing problem. Binary numbers do not perfectly match because of the production of redundant positions. It has been solved justifiably with 27 ternary numbers which exactly fits 20 standard and 5 non-standard amino acids with GAP and UNKNOWN. One of the best encodings is done in order of using molecular weight of amino acids into ternary numbers. We have formulated one mathematical formula "Impression" which is applied on ternary numbers. Based on this parameter it has been clearly explained the degeneracy of codon table. Further, on using graph theoretic model we have shown the co-relation of existing classes with our classification result for Iron protein family. On using our classification results some proteins are clubbed together into a single class from different existing classes which may be used for new drug design.